\documentclass [dvips, 12pt] {article} 
\usepackage {graphicx}
\usepackage {pifont}
\newcommand {\pion} {\ensuremath {{\pi^0}} }

\newcommand*{\meson}[1]{\ensuremath {#1^\pm}}

\newcommand {\cd} {\normalsize \textcircled {\em \scriptsize d}}
\newcommand {\cu} {\normalsize \textcircled {\em \footnotesize u}}
\newcommand {\cs} {\normalsize \textcircled {\em \footnotesize s}}
\newcommand {\cc} {\normalsize \textcircled {\em \footnotesize c}}
\newcommand {\cb} {\normalsize \textcircled {\em \footnotesize b}}

\newcommand {\cdb} {\normalsize \textcircled {\em \scriptsize   $\bar{d}$}}
\newcommand {\cub} {\normalsize \textcircled {\em \footnotesize $\bar{u}$}}
\newcommand {\csb} {\normalsize \textcircled {\em \footnotesize $\bar{s}$}}
\newcommand {\ccb} {\normalsize \textcircled {\em \footnotesize $\bar{c}$}}
\newcommand {\cbb} {\normalsize \textcircled {\em \scriptsize $\bar{b}$}}

\begin {document}

\begin {center}
{\Large Precise mass spectrum of mesons with open charm in the harmonic quarks
 and oscillators. }\par\bigskip 
{\large Oleg~A.~Teplov}\par\smallskip 
Institute of metallurgy and materiology of the Russian Academy of Science, 
Moscow.
\par e-mail: teplov@ultra.imet.ac.ru 
\end {center}

\begin{abstract}

The harmonic quarks and their complete oscillators are presenting the 
unprecedented exact solution for the mass spectrum
 of mesons with an explicit charm. The experimental 
and calculated spectrums coincide with standard deviation in 1.8 MeV.
The four doublets and experimental errors of meson masses give the main 
contribution in this value. The harmonic spectrum is not  
sensitive  to an isospin. The spectrum is simple, complete, 
precise and logically clear. It's easy to make predictions. 
The spectrum has not free parameters. It's supposed that the spectrum 
of harmonic and mixed levels is an exterior spectrum 
of allowed energy states in relation to a QCD states. 
The features of the harmonic spectrum, some levels of charm hadrons,  
the acceptor property of a $c$-quark, the mass rank of quarks and 
the nature of leptons are discussed.
 
\end{abstract}

\section {Introduction}

  In the present publication the harmonic quarks and their complete 
oscillators are used for decoding mass spectrum of explicitly 
charmed mesons.
      In~\cite{my1, my2} regularity in a spectrum of meson masses 
was detected and the simple harmonic model of quarks was offered.
 Furthermore their masses calculated with precision about 0.005\%. 
In the same works it is shown, that the harmonic model and the harmonic
 quarks are capable to explain experimental masses of some particles,
 including both a muon, and a neutral pion. Since article~\cite{my3},
 we began a systematic application of the harmonic quarks and their
{\em complete harmonic oscillators} (afterwards also referred to as 
{\em harmonic oscillators}) for exposition of a hadron mass spectrum. 
In~\cite{my3} 
was shown, that the masses of light hadrons up to 1000 MeV are arranged
 nearly or directly in potential wells of complete harmonic oscillators
 and their combinations.  
Here, in this paper, we shall use the same notation, as in~\cite{my3},
 namely: standard symbols ($u, s, c$ …) shall designate
 {\em the harmonic quarks}, while encircled symbols shall designate quarks,
 which are bound in the complete oscillator or are having the same energy 
 as in oscillator. Besides, a simple and combined potential
 wells we shall frequently call {\em harmonic levels} or simply 
levels.

   In~\cite{my3} was shown, that there are no light and strange mesons with
the mass less than their respective oscillator. In other words, the energy
 of harmonic oscillator is a lower limit for ground state of hadrons
 with the given flavor. 
 Thus, mass of \pion is greater by 0.7 MeV than oscillator \cu\cub, and 
mass of \meson{K} is greater by 2.3 MeV than oscillator \cs\csb. 
For mesons with an explicit charm this statement is also valid. 
The masses of quarks,
 harmonic oscillators and some mesons with corresponding flavors 
 are given in table 1.

\begin{center}
Table 1. The masses of harmonic quarks and their complete oscillators.  

  \medskip\small 
  \begin{tabular}{|c|c|c|c|c|c|}
    \hline
Quark & Quark mass, & Harmonic & Oscillator energy,& Meson & Mass of meson, \\
       &  MeV & oscillator  & MeV & & MeV \\
    \hline
   $d$ & 28.8106 & \cd \cdb  & 36.683 & - & - \\
\hline
   $u$ & 105.441 & $\cu \cub$ & 134.251 & \pion     & 134.98 \\
\hline
   $s$ & 385.891 & $\cs \csb$ & 491.332 & \meson{K} & 493.65 \\
\hline
   $c$ & 1412.28  & $\cc \ccb$ & 1798.17  & $D^0$     & 1864.6 \\
\hline
   $b$ & 5168.7   & \cb \cbb & 6581.0   & $B_c$     & 6400   \\
\hline

  \end{tabular}

\end{center}

\section {The spectrum of harmonic levels } 

The mass of the first $c$-hadron $D^0$
 is located at 66.4 MeV above an oscillator \cc\ccb. We can note that 
the $c$-quark needs some additional energy for formation of $D^0$, 
in comparison with the first mesons for $u$- or $s$-flavor (tab.1).
    At study of a mass spectrum of explicitly charmed mesons 
we shall not analyze dispersion curves as in~\cite{my3}, 
but start instead with immediate interpretation
 of the experimental masses of 
 mesons~\cite{parti}--\cite{selex}.
Unlike the light mesons the harmonic oscillator \cc\ccb\ (i.e. main base level) of the charm mesons has large energy (1798.17 MeV).
 It considerably facilitates the analysis of their 
spectrum of masses, since mass differences between the mesons 
and \cc\ccb\, oscillator are less than 850 MeV.  These differences 
are given in table 2. The charged mesons of doublets are omitted.
The majority of these differences can be explained lightly with use of 
 oscillators (\cs\csb, \cu\cub, \cd\cdb) and quark-antiquark pairs
 ($u\bar{u}$, $s\bar{s}$).

 $D^*$(2290)$^0$,  $D_1$(2420)$^0$  and  $D_2$(2460)$^0$ form a series
 on the basis 
of oscillators \cc\ccb\, + \cs\csb.

\begin {figure} [htb]
\begin {center}
\includegraphics [scale =.7] {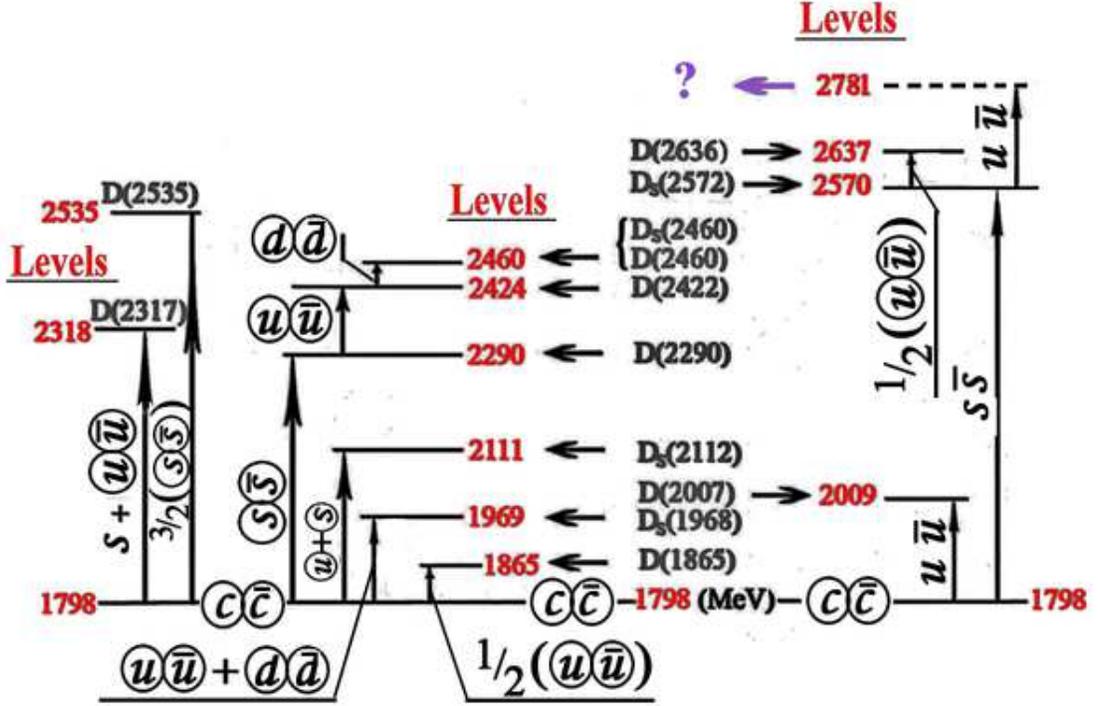}
\par
\caption [] {The complete spectrum of harmonic and mixed levels 
for mesons with open charm.
\label {fig:spec}}
\end {center}
\end {figure}

$D_s^\pm$, $D^*$(2007)$^0$ 
and $D_{s2}(2573)^\pm$ are also arranged in simple potential wells.
We can see that this levels is very simple and 
in the same time it surprisingly  exactly agrees with masses of mesons 
with open charm. In table 2 and fig.~\ref{fig:spec} are shown the levels 
which are obtained exclusively from quark-antiquark oscillators
 with various flavors. 
There is the unique series of  three levels which are formed 
by consecutive addition of other  oscillators to \cc\ccb-oscillator: 

\cs\csb; \cs\csb\, + \cu\cub; \cs\csb\, + \cu\cub\, + \cd\cdb.
  
Thus, this is purely harmonic series. The last level is the total energy
of all four complete harmonic oscillators of quarks $c, s, u, d$. 
These levels are filled by doublets $D^*$(2290)$^0$, $D_1$(2420) 
and $D_2$(2460).
{      
\begin{center}
Table 2. Additional energy of mesons above an oscillator \cc\ccb. 
  \medskip\small 
  \begin{tabular}{|c|c|c|c|c|}
    \hline
 Mesons & Meson masses, & Additional energy, & Oscillators or & Energy of quark \\
       &  MeV & MeV &quark group& group, MeV \\
    \hline
 $D^0$   &1864.6  &66.4  & - & -  \\
\hline
 $D_s^\pm$   &1968.3  &170.1  &\cu\cub\, + \cd\cdb &170.93 \\
\hline
 $D^*$(2007)$^0$  &2006.7  &208.5  &$u\bar{u}$   &210.88  \\
\hline
$D_s^{*\pm}$   &2112.1  &313.9  & -  &  -  \\
\hline
$D^*$(2290)$^0$  &2290  &491.8  &\cs\csb &491.3  \\
\hline
$D_{sJ}$(2317)$^\pm$ &2317.4  &519.2  & - & -  \\
\hline
 $D_1$(2420)$^0$ &2422.2  &624.0  &\cs\csb\, + \cu\cub  &625.58 \\
\hline
 $D_2^*$(2460)$^0$ &2458.9  &661.2  &\cs\csb+\cu\cub+\cd\cdb  &662.26  \\
\hline
$D_{sJ}$(2460)$^\pm$ &2459.3  &661.1  &\cs\csb+\cu\cub+\cd\cdb  &662.26    \\
\hline
$D_{s1}$(2536)$^\pm$  &2535.35  &737.2  & -  & - \\
\hline
$D_{s2}$(2573)$^\pm$ &2572.4  &774.2  &$s\bar{s}$  &771.78   \\
\hline
 $D^*$(2640)$^\pm$ &2637  &838.8  & - & - \\
\hline

  \end{tabular}

\end{center}
}

\par
\par

 Furthermore, the level 2460 generates the strange meson 
$D_{sJ}$(2460)$^\pm$. 
Long-lived meson $D_s^\pm$ is filled also the purely harmonic level 
from three oscillators
 \cc\ccb\, + \cu\cub\, + \cd\cdb. Unlike the level for $D_{sJ}$(2460)$^\pm$,
 the level
 for $D_s^\pm$ does not contain \cs\csb\-oscillator, but it is purely
 harmonic level and can be included the above named series of levels
 as its fourth member. 
    In the right part of fig.~\ref{fig:spec} two hybrid levels are presented. They are
 formed by excitation of quark - antiquark
 pairs $u\bar{u}$ and $s\bar{s}$ above the base level \cc\ccb. 
 A distinctive feature of these levels is the absence 
 of excited levels with  pair $d\bar{d}$. 
 
After this simple procedure there are remained only 5 mesons without
 interpretation of additional energy:

 $D^0$,$D_s(2112)^\pm$, 
 $D_{sJ}$(2317)$^\pm$, $D_{s1}(2536)^\pm$ and $D^*$(2640)$^\pm$.

First of all we shall note that the mass of $D^*$(2640)$^\pm$ is precisely equal
to the mass of $D^0$ with the additional excitation $s\bar{s}$. The mass 
of $D^0$ is greater  by 66.4 MeV than a base level. This value
 is practically equal to the energy of a $u$-quark in its own oscillator
 \cu\cub, i.e. half the energy of an oscillator (67.1 MeV). Then, 
the potential well for $D^0$ can be written as: \cc\ccb\, + \cu. 
However the author consider the following form is more right (discussion 
of this problem will follow): \cc\ccb\, + 1/2(\cu\cub). Hence the well 
for $D^*$(2640)$^\pm$
 is: \cc\ccb\, + $s\bar{s}$ + 1/2(\cu\cub).

There are only 3 states with a strangeness to be considered. 
 We use the same tactics as for $D^0$. We assume 
that half energy from an oscillator \cs\csb\, is the energy of  
the harmonic bound $s$-quark (245.7 MeV). 
We can easily gain levels for $D_s^*$(2112)$^\pm$ and $D_{s1}$(2536)$^\pm$:

\cc\ccb\, + 1/2(\cu\cub\, + \cs\csb) and \cc\ccb\, + 3/2(\cs\csb)

with energies 2111 and 2535 MeV respectively.
The potential well for the last particle $D_{sJ}(2317)^\pm$ seems a little
 debatable: $c + ss$ + \cu\cub\, with energy of a level 2318 MeV.
 The level can also be written as: \cc\ccb\, + $s$ + \cu\cub. 
 The supposed
 configuration of level $D_{sJ}(2317)^\pm$ is a little unusual: it contains 
an odd quark combination in a configuration of the level though a similar
 combinations are present in configurations of levels $D_{s1}$(2536)$^\pm$, 
$D^0$ and $D^*$(2640)$^\pm$.
The results of the full decoding of the meson levels
 with open charm are presented in table 3 and fig.~\ref{fig:spec}.
The dotted line in fig.~\ref{fig:spec} shows one of the guessed levels.
 The number of them can be easily increased, but nevertheless 
these levels are only guessed levels. We are just starting to study
both the reasons of their generation and real laws
which operate in this area.
Model spectrum is calculated from a uniform principle 
without the free parameters. The standard deviation of charm mesons 
from their calculated levels is equal to 1.8 MeV. 
The calculation was made for 16 mesons. 
The four doublets and experimental errors of meson masses 
give the main contribution in this value.
So root-mean-square experimental error 
for 15 mesons($D$(2290) is excluded) is equal to 2.1 MeV. 

{
\begin{center}
Table 3. The complete interpretation of meson spectrum with open charm
 and calculated energy of their levels.  
  \medskip\small 
  \begin{tabular}{|c|c|c|c|c|c|}
    \hline
 Mesons & Meson masses, & Rest of mass & Oscillator group & Energy &Calculated \\
                &  MeV &  above \cc\ccb, &or/and& of group, & energy of \\
                &      & MeV               &quark group           &  MeV& level, MeV \\
    \hline
 $D^0$   &1864.6  &66.4  &1/2(\cu\cub) & 67.1 &1865.3 \\
\hline
 $D_s^\pm$   &1968.3  &170.1  &\cu\cub\, + \cd\cdb &170.93&1969.1 \\
\hline
 $D^*$(2007)$^0$  &2006.7  &208.5  &$u\bar{u}$   &210.88&2009.1  \\
\hline
$D_s^{*\pm}$   &2112.1  &313.9  &1/2(\cs\csb\, + \cu\cub) &312.79&2111.0  \\
\hline
$D^*$(2290)$^0$  &2290  &491.8  &\cs\csb &491.3&2289.5  \\
\hline
$D_{sJ}$(2317)$^\pm$ &2317.4  &519.2  & s + \cu\cub &520.14 &2318.3 \\
\hline
 $D_1$(2420)$^0$ &2422.2  &624.0  &\cs\csb\, + \cu\cub  &625.58&2423.8 \\
\hline
 $D_2^*$(2460)$^0$ &2458.9  &661.2  &\cs\csb+\cu\cub+\cd\cdb  &662.26 &2460.4  \\
\hline
$D_{sJ}$(2460)$^\pm$ &2459.3  &661.1  &\cs\csb+\cu\cub+\cd\cdb  &662.26 &2460.4   \\
\hline
$D_{s1}$(2536)$^\pm$  &2535.35  &737.2  & 3/2(\cs\csb) &	737.00 &2535.2 \\
\hline
$D_{s2}$(2573)$^\pm$ &2572.4  &774.2  &$s\bar{s}$  &771.78 &2570.0  \\
\hline
 $D^*$(2640)$^\pm$ &2637  &838.8  &$s\bar{s}$ + 1/2(\cu\cub) & 838.91 &2637.1 \\
\hline

  \end{tabular}

\end{center}
}

\section {DISCUSSION OF RESULTS}

     Simple, logical, complete and precise interpretation of mass
 spectrum  of mesons with open charm demonstrates to us that {\em harmonic 
quarks are a new true message from a microcosm}. At present the 
particle physics has not other quantitative theory or a model
 which achieve such exact results. Now the question - To be or not to be? - 
in relation of the harmonic quarks can be closed. They exist and their
 rest masses are calculated precisely enough. Actually there 
are another problems. How should the harmonic 
quarks be integrated correctly with QCD and Standard Model? 

How may we create field theory in which the mass ratio 
of neighboring quarks is a constant equal to $\pi/(4-\pi)$ and 
which has a solution with the bound harmonic  states 
of the  quark-antiquark pairs? 
What should a boundary conditions be at low and high energies 
to restrict a theoretical quark spectrum to only 
really observable quarks?
Why are the complete harmonic oscillators so important
 for hadronic physics?

\subsection {Some features of model spectrum}

\begin {figure} [htb]
\begin {center}
\includegraphics [scale =.6] {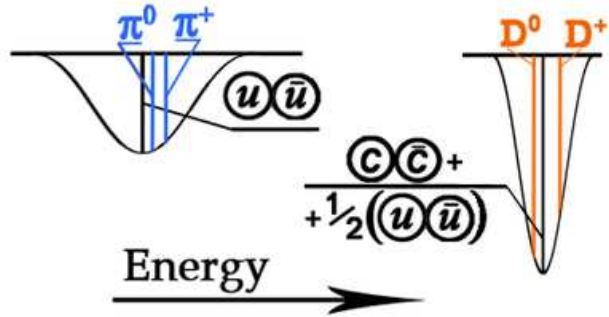}
\par
\caption [] {The scheme of harmonic wells and arrangement of hadron doublets.
\label {fig:schem}}
\end {center}
\end {figure}

Returning to the subject of the paper, we should note one 
important observed fact. It is necessary to distinguish quark
 interpretation of harmonic level and an actual quark composition
 of a hadron on this level. It is especially important in regard to 
 long-lived particles. The simple example is any multiplet 
which fills only one harmonic level. 
{ \em The harmonic levels are indifferent to an isospin.} So, for each pair
 of mesons ${K^{*\pm}}$ and $K^{*0}$~\cite{my3} or $D^0$ and $\meson{D}$ there 
is the harmonic level
 which has the position between a mesons of doublet. Energy of level
 corresponds to centre of potential well as it is shown in 
fig.~\ref{fig:schem}.
 The multiplets can occupy various positions in these wells.

\par
\par

 For example, pions and kaons have a little shift from centre in side of energy increase.
 These effects can be considered in detail in tables of this work 
and~\cite{my1, my2, my3}. It is necessary to notice, a mesons
 with different secondary flavor, for example $D_2^*$(2460) 
 and $D_{s1}$(2460)$^\pm$,
can be arranged on one harmonic level. 
{\em Probably, the harmonic levels are also not sensitive to secondary 
flavor.}

\begin {figure} [htb]
\begin {center}
\includegraphics [scale =.6] {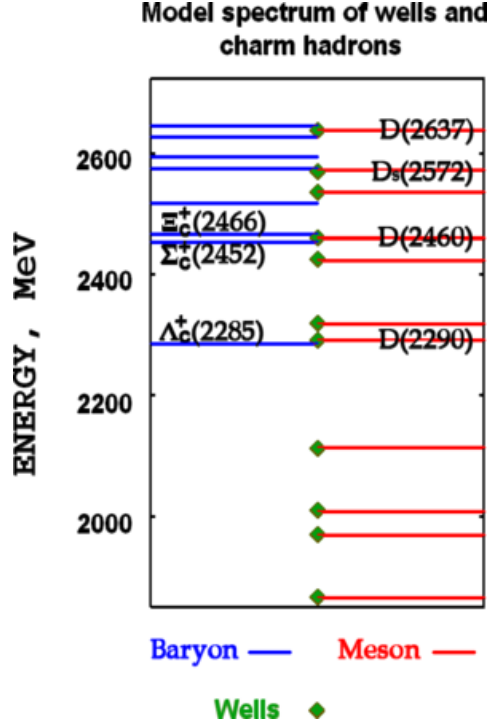}
\par
\caption [] {The harmonic levels and their hadrons
with open charm. \label {fig:arrang}}
\end {center}
\end {figure}

 In table 4 the most important potential wells are given
 for 4 flavors and there are the hadrons which perhaps connected
 with these wells.
The all levels of a charmed meson spectrum contains the ground
 oscillator - \cc\ccb. Depth of a potential well of an oscillator 
is proportional to its mass. For \cc\ccb\, the depth of the well is 1026 MeV,
 if it is counted from a rest energy of quark-antiquark $c$-pair. 
It is logical to assume, that the potential well of an oscillator
 is capable to keep more light quarks and oscillators.
 \cc\ccb-well is enough deep and it is capable 
to keep $d$-, $u$- and $s$-quarks as well as their oscillators. It actually 
confirms the obtained spectrum, especially a series of harmonic levels.
In table 4, besides a series on the basis of \cc\ccb, we can observe 
other series.  They are built on deep wells from 
the combinations of multiple $s$-oscillator: 2(\cs\csb) and 3(\cs\csb). 
We can see that the harmonic wells can be occupied not only
 by mesons, but also baryons. 
Thus we come to the conclusion, that {\em probably, the harmonic potential
 wells are also not sensitive to type of a hadron}. They can be 
a manifestation of the more common rules and they determine
 energy spectrum of all hadrons.
On fig.~\ref{fig:arrang} the mass spectrum of charm baryons is added to the data 
obtained in this work. It is obvious that the lightest baryons
 ($\Lambda_c^+$, $\Sigma_c^+$, $\Xi_c^+$) are also close 
to the found harmonic levels.  

Hence, the spectrum of harmonic levels (at least for the charmed
 mesons) may be determined as the spectrum of allowed states.
There is an initial understanding that hadronic states of a QCD are limited 
by energy spectrum of harmonic levels. The hadronic states 
are perhaps forced to the nearest harmonic potential wells
 or mixed wells.    

\begin{center}
Table 4. The most important potential wells from complete harmonic  
 oscillators \\ of quarks: $d, u, s, c$.
  \medskip\small 
  \begin{tabular}{|c|c|c|c|}
    \hline
Harmonic potential &The total energy of & Hadron mass, & Hadron \\
wells       &well oscillators, MeV &  MeV~\cite{parti, belle} &  \\
    \hline
 \cu\cub  & 134.25 & 134.98 & \pion \\
\hline
  \cs\csb & 491.33 & 493.65 & \meson{K} \\
\hline
  \cs\csb\, + \cu\cub &625.6 & - & - \\
\hline
 3\cs\, + 3\cu & 938.37&	938.27 & $p,~\bar{p}$ \\
\hline
 \cs\csb\, + 3(\cu\cub) & 894.1 & 891.7  & $K^{*\pm}$ \\
    & & 896.1 & $K^{*0}$ \\
\hline
2(\cs\csb)  & 982.66 & 984.8 & $a_0$(980) \\
    & & 980 & $f_0$(980) \\
\hline
  2(\cs\csb) + \cd\cdb & 1019.34 & 1019.46 & $\phi$(1019) \\
\hline
  2(\cs\csb) + \cu\cub & 1116.9 & 1115.7 & $\Lambda(1115)$ \\
\hline
 2(\cs\csb) + 3(\cu\cub) & 1385.4 & 1383.7  & $\Sigma(1385)$ \\
\hline
  3(\cs\csb) & 1474.0 &	~1474 & $a_0$(1450) \\
  &  &	~1476 & $\eta(1475)$ \\
  &  &	~1465 & $\rho(1450)$ \\
\hline
  3(\cs\csb) + \cd\cdb & 1510.7 & 1507 & $f_0$(1500) \\
 \hline
  3(\cs\csb) + \cu\cub & 1608.2 & ~1600 & $\Lambda(1600)$ \\
   &  &	~1596 & $\pi_1$(1600) \\
 \hline
 3(\cs\csb) + 3(\cu\cub) & 1876.8 & 1876.7  & $p\bar{p}$ \\
\hline

\cc\ccb\, + 1/2(\cu\cub)  & 1865.3 & 1864.6 & $D^0$ \\
\hline
\cc\ccb\, + \cu\cub\, + \cd\cdb & 1969.1 & 1968.3 & $D_s^\pm$ \\
\hline
\cc\ccb\, + \cs\csb & 2289.5 & 2290 & $D^*$(2290)$^0$ \\
\hline
\cc\ccb\, + \cs\csb\, + \cu\cub & 2423.8 & 2422.2 & $D_1$(2420)$^0$ \\
\hline
\cc\ccb+\cs\csb+\cu\cub+\cd\cdb & 2460.4 & 2458.9  & $D_2^*$(2460)$^0$ \\
 & & 2459.3 & $D_{s1}$(2460)$^\pm$ \\
\hline

  \end{tabular}

\end{center}

\par
\par

\subsection {A feature levels}
Let's now to discuss an individual features of some levels. 
The spectrum has a potential wells with an odd number of quarks: 

\cc\ccb\, + 1/2(\cu\cub) for $D^0$ and \meson{D};    
\cc\ccb\, + \cu\cub\, + $s$ for $D_{sJ}$(2317)$^\pm$;

\cc\ccb\, + 3/2(\cs\csb) for $D_{s1}$(2536)$^\pm$;
\cc\ccb\, + $s\bar{s}$ + 1/2(\cu\cub) for $D^*$(2640)$^\pm$.

The simplest variant for an explanation of this phenomenon
 can be some pair solutions for particles
 and antiparticles, for example, as the similar solution 
 for a pair $e^+e^-$. 
This pair solution, i.e. doubling, ensures us even configurations.
 So, level $D^0\bar{D}^0$ may be written as: 2(\cc\ccb) + \cu\cub. 
This configuration is completely similar to configuration of $\phi$(1019) 
in tab.4. They are differed from each other only by shift on one step 
on a quark scale.
 However, this shift changes a situation, since energy and a numbers of
 quarks with charges $\pm1/3e$ and $\pm2/3e$ are different. The quarks
 with a charge $\pm1/3e$ for $\phi$-level are completely replaced with the quarks with a charge $\pm2/3e$. 
We have not the complete physical analogy of two levels
 and therefore the different ending is possible: level $\phi$ keeps up
 whereas level $D^0\bar{D}^0$ breaks up. 

The decay of the unique level 6\cs\, + 6\cu\, on $p\bar{p}$~\cite{my3} 
is also given in table 4. 

Other explanation for a level \cc\ccb\, + 1/2(\cu\cub) can be proposed
with the absence of a level \cc\ccb\, + \cu\cub. 
Perhaps, it is forbidden for still unknown reasons. The basic level \cc\ccb\, is also forbidden. The level 
\cc\ccb\, + 1/2(\cu\cub)
 is arranged in the middle between forbidden levels, and  
it is probably a degenerate level.       

\subsection {The mass rank of quarks} 

We have two levels which contain a half $u$-oscillator 1/2(\cu\cub), 
i.e. harmoniously
  bound $u$-quark, similar to $u$-quarks in a neutral pion.
 A pion in $u\bar{u}$-phase (this restriction follow of a QCD with its
 a superposition of states $u\bar{u}$ and $d\bar{d}$) is practically 
pure harmonic state of a bound $u\bar{u}$-pair.
It is the lowermost level in a spectrum of hadrons and only therefore
 it is unique. It is a single long-lived meson and a truly neutral
 particle is directly arranged in a potential well \cu\cub.
 There is not any other truly neutral particle, which is arranged in  potential well of kind 
${\textcircled {\em \footnotesize q}}{\textcircled {\em \footnotesize $\bar{q}$}}$. 
So, $K^0$ has an antiparticle
 with an antiquark composition. In a state of $u$-quarks 
the pion can be written down as: \cu\, + \cub,

 where \cu\,is a bound $u$-quark with the rest mass equal to 
1/2 of energy of a $u$-oscillator, i.e. 67.1257 MeV.
Then the dynamic energy of \cu-quarks in pion is  $m_\pion$ - 134.2514
 and equal to 0.7252 MeV.
Together with a Coulomb electrostatic energy of \cu\cub-pair
it will correspond to the kinetic energy of \cu-quarks on small distances,
 where the color energy is small enough.
 Accepting this structure of \pion, we actually assume, 
that the rest mass 
of a quark in harmonic bound state decreases in $\pi$/2 times.

This phenomenon possible to name as {\em diminution of the mass rank
 of quark} (DMRQ).

 In essence, all levels (potential wells) 
on fig.~\ref{fig:spec}
 are obliged by their existence to DMRQ. 
Even the levels for particles $D^*(2007)^0$ and $D_{s2}$(2573)$^\pm$ 
are constructed 
on basis of $c$-quarks with DMRQ. We can see that light quarks can decrease 
their mass rank in the strong field of a heavy quark or its oscillator. 
We can even note, a quarks in state DMRQ of first order can again form 
a harmonic oscillator with the further decrease of mass. 
There may to occur a secondary DMRQ (see a level for $D_s^*$(2112)$^\pm$). 

In the same time the spectrum contain levels with odd number of \cu\, 
quarks. 
However the majority of levels consist from neutral groups 
with even number of quarks. Hence for $u$-quark we can suppose 
an existence of his neutral state of kind $u^0$ and $\cu^0$.
It is possible to assume that the charge 2/3$e$ of $u$-quark may be
 not elementary in comparing with 1/3$e$ of other quarks.
 Then a $u$-quark contains two elementary charges. 
Therefore it is possible to imagine the reaction 
of a part annihilation of $u\bar{u}$ pair which goes to formation of $u_0$,
 and even to $\cu^0$ in a field of a heavy quark in result of DMRQ.

\subsection {Acceptor property of $c$-quark}

The levels \cc\ccb\, + 1/2(\cu\cub)\, and \cc\ccb\, + $u\bar{u}$ are allowed. 
There is an electromagnetic
 transition between mesons of these levels. The quark reaction 
of transition
 from the first level to second level can be written as follows:

$u$ + $\bar{u}$ = ($u$/2 + $u$/2) + ($\bar{u}$/2 + $\bar{u}$/2) $\Rightarrow$ 1/2(\cu\cub) = $\cu^0$.

In this reaction we assume that $u$-quark consists from two parts
 with charges equal 1/3$e$ as well as any other quark 
with a charge 2/3$e$. 
It is supposed, that the powerful field of an  oscillator \cc\ccb\, promotes
 this reaction, i.e. a level \cc\ccb\, + 1/2(\cu\cub),
 can be written as:  (\cc\, + \cu/2) + (\ccb\, + \cub/2).
 The last record could mean that the $c$-quark has very
 unsaturated internal structure, and it is capable to accept an additional 
energy. {\em A $c$-quark is an acceptor. A $b$-quark is a donor.} 

  A $c$-quark has a certain acceptor properties. Only the spectrum 
of $b$-hadrons is arranged below  \cb\cbb-oscillator. The mass of \meson{b} 
 is only greater by 5 MeV than total mass of $u$- and $b$-quarks,
 and is 1302 MeV less than \cb\cbb-oscillator. The mass 
of a $D^0$ is greater by 344 MeV than the total mass of $u$- and $c$-quarks 
and it is greater by 66.4 MeV than \cc\ccb-oscillator. If the tendency
 fulfilled for a $b$-quark, the additional energy (energy over masses
 of $u$- and $b$-quarks) would be about 1306 MeV. However, for existence 
of a $b$-meson an additional energy is not required, 
i.e. {\em the $b$-quark
 is not an acceptor}. On the contrary, the heavy $b$-quark,
 being the owner of a large potential energy, can even have
 donor properties. So, the mass  of $B_s$ is less than the total mass
 of $s$- and $b$-quarks. It is possible that the nature of acceptor 
or donor properties of quarks is connected with difference 
of internal structure of quarks.
This difference is reflected in their charges (1/3$e$ or 2/3$e$).
 Quarks with the charge 1/3$e$ is perhaps 
 more simple objects. Besides it is natural that 
this distinction of energy properties are more clearly expressed 
at heavy quarks.
We shall return to this problem, but at present time
 it is enough to formulate the presence of a problem.

\subsection {The $\tau$-lepton}

\begin {figure} [htb]
\begin {center}
\includegraphics [scale =.7] {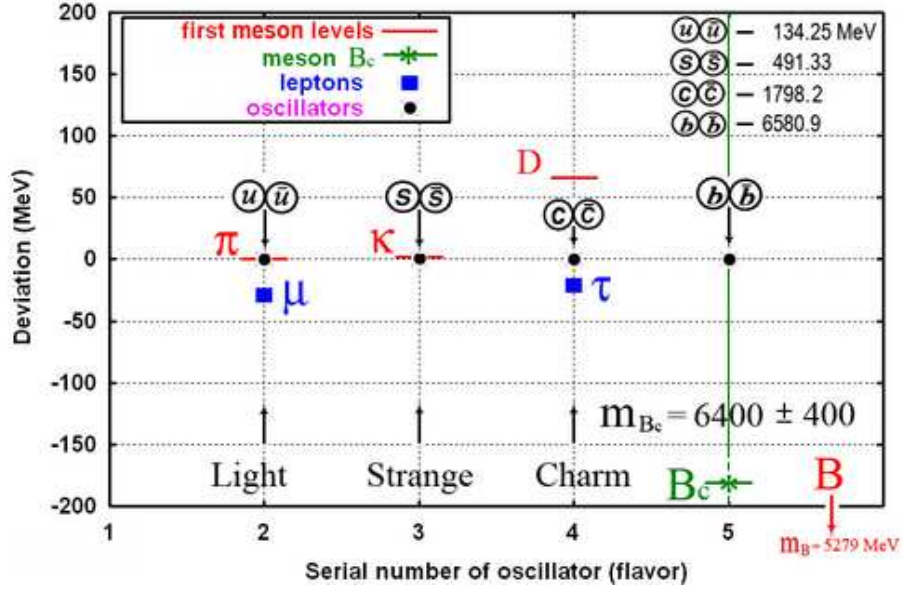}
\par
\caption [] {The deviations of first flavor mesons and leptons from 
their harmonic oscillators.  \label {fig:flavors}}
\end {center}
\end {figure}

The mass of $\tau$-lepton is 1777 MeV. It is only 21 MeV less 
than the mass of \cc\ccb-oscillator, 
but this also means, that birth of  meson
 with a rest mass less than  \cc\ccb-oscillator is forbidden. In~\cite{my2}
 we assumed, that if the muon is formed on the mass
 basis of the $u$-quark then the $\tau$-lepton may be also formed 
on the mass basis of the $c$-quark. 
A creation of leptons is accompanied by inhibition or annihilation 
of color charge and by increase of electrical charge by 1/3$e$. 
It is important to note that masses of both leptons are only 
on 20-30 MeV less than 
the masses of corresponding harmonic oscillators (see fig.~\ref{fig:flavors}).

\par
\par

The  harmonic oscillators on fig.~\ref{fig:flavors} are located on zero line.

 The other validity 
of this important assumption proves the fact of approximate equal lifetimes 
of quarks and corresponding leptons.
So, the lifetime of $\tau$-lepton is 3*$10^{-13}$ sec. This value is practically 
the same as the lifetimes of $D^0$ and $D_s^\pm$ mesons. 
Lifetime of these mesons 
is defined by the lifetime of a $c$-quark.
 Besides, both leptons are formed on a mass basis
 of quarks with a charge 2/3e. If it is quite clearly for muon
(the masses of muon and $u$-quark are practically equal, even 
the distinction of masses in 0.2 MeV has quite clear 
interpretation~\cite{my1, my2}),
 for $\tau$-lepton such obviousness is absent. However the author 
believes that all is true, therefore it also illustrates an acceptor
 properties of $c$-quark.

\section {Conclusion}
\begin {itemize}

\item The precise model spectrum of  mesons with open charm 
         is formed exclusively on rest masses of the harmonic quarks 
         and their complete oscillators. The spectrum has not 
         free parameters.
	 It is calculated exclusively with use of hard harmonic quark model.   
\item It is assumed the harmonic spectrum is an allowed energetic states
         in relation to a QCD states.  
\item The harmonic quarks and their complete oscillators 
         are actual objects of a microcosm.
\item A rest masses of the harmonic quarks are calculated 
         with the declared precision $\sim$~0.005\%.
\item The harmonic quarks can decrease their mass rank at certain conditions. 
\item The complete harmonic oscillators are the basis of the strong
         coupling and physics of hadrons.
\item The $c$-quark has acceptor properties. 
\item It is obtained additional arguments in support of assumption that
  the $\tau$-leptons are formed with use of $c$-quarks.

\end {itemize}

\end {document}